\documentclass[prd,reprint,onecolumn,showpacs,preprintnumbers,nofootinbib]{revtex4-1}
\usepackage{hyperref}
\usepackage{latexsym,epsfig,graphicx,amsmath,amssymb}

\def\de{\partial}

{\rm }

\def\k{\kappa}

\def\be{\begin{equation}}
 \def\ee{\end{equation}}
 \def\bea{\begin{eqnarray}}
 \def\eea{\end{eqnarray}}

\newcommand{\fr}{\frac}
\newcommand{\pr}{\prime}

\def\2{\frac{1}{2}}
\def\4{\frac{1}{4}}

\def\nn{\nonumber}

\def\PR#1{{Phys.\ Rev.\ D \bf #1}}
\def\PRL#1{{Phys.\ Rev.\ Lett.\ \bf #1}}

\catcode`\@=11

\def\@normalsize{\@setsize\normalsize{15pt}\xiipt\@xiipt
\abovedisplayskip 14pt plus3pt minus3pt%
\belowdisplayskip \abovedisplayskip
\abovedisplayshortskip  \z@ plus3pt%
\belowdisplayshortskip  7pt plus3.5pt minus0pt}
\def\small{\@setsize\small{13.6pt}\xipt\@xipt
\abovedisplayskip 13pt plus3pt minus3pt%
\belowdisplayskip \abovedisplayskip
\abovedisplayshortskip  \z@ plus3pt%
\belowdisplayshortskip  7pt plus3.5pt minus0pt
\def\@listi{\parsep 4.5pt plus 2pt minus 1pt
            \itemsep \parsep
            \topsep 9pt plus 3pt minus 3pt}}

\def\underline#1{\relax\ifmmode\@@underline#1\else
        $\@@underline{\hbox{#1}}$\relax\fi}
\@twosidetrue \relax

\catcode`@=12

\catcode`\@=11

\def\section{\@startsection{section}{1}{\z@}{3.5ex plus 1ex minus
   .2ex}{2.3ex plus .2ex}{\large\bf}}

\def\ps@headings{\def\@oddfoot{}\def\@evenfoot{}
\def\@oddhead{\hbox{}\hfill
        \makebox[.5\textwidth]{\raggedright\ignorespaces --\thepage{}--
        \hfill }}
\def\@evenhead{\@oddhead}
\def\subsectionmark##1{\markboth{##1}{}}
}

\ps@headings

\catcode`\@=12

\begin{document}

%
%


%

\title{Phase Transition to a Hairy  Black Hole in Asymptotically Flat Spacetime}

\author{Theodoros Kolyvaris}\email{teokolyv@central.ntua.gr}
\author{George Koutsoumbas}\email{kutsubas@central.ntua.gr}
\author{Eleftherios Papantonopoulos}\email{lpapa@central.ntua.gr}
 \vspace{.2in}
\affiliation{Department of Physics, National Technical University of
Athens,
Zografou Campus GR 157 73, Athens, Greece.}
\author{George Siopsis}\email{siopsis@tennessee.edu}
\affiliation{Department of Physics and Astronomy, The
University of Tennessee, Knoxville, TN 37996 - 1200, USA.}

\date{August 2013}

\begin{abstract}

We discuss a phase transition of a Reissner-Nordstr\"om black hole to a hairy black hole in asymptotically flat spacetime. The hair is due to a massive charged scalar field. The no-hair theorem is evaded thanks to a derivative coupling of the scalar field to the Einstein tensor. The resulting hairy configuration is spherically symmetric. We solve the equations analytically near the transition temperature and show that the hair is concentrated near the horizon decaying exponentially away from it.

\end{abstract}

\maketitle

\section{Introduction}

Scalar-tensor theories belong to a class of theories that modify
the Einstein's theory of gravity and have been under intense
investigation over the last few years.  The most interesting class of
scalar-tensor models  are described by the Horndeski Lagrangian
\cite{horny}, which gives second-order field equations in four
dimensions. It was soon realized that these scalar-tensor models
of modified gravity share a classical Galilean symmetry
\cite{Nicolis:2008in, Deffayet:2009wt, Deffayet:2009mn}, and most
of these models represent a special case of Horndeski's theory.

One of the terms appearing in the Horndeski Lagrangian is the
derivative coupling of a scalar field to Einstein tensor. The
cosmological implications of the derivative coupling to gravity
was first discussed in \cite{Amendola:1993uh}, and subsequently it
was shown that the presence of this coupling in the Lagrangian
gives second-order field equations  \cite{Sushkov:2009hk}, in
accordance with Horndeski's theory. A cosmological model was
discussed in \cite{Gao:2010vr,Sushkov:2012za}, where the derivative
coupling was introduced to explain  early as well as  late expansion of
the Universe. Quintessence and phantom cosmology  in the presence of this
coupling was presented in
\cite{Saridakis:2010mf,Granda:2012tx}, while accelerating expansion was discussed in
\cite{Granda:2009fh,Gubitosi:2011sg}. The early inflationary phase
was studied in \cite{Germani:2010gm}, where it was found that the
derivative coupling acted like a friction term. It was also
found that this term had a Galilean symmetry
\cite{Germani:2011bc}.

Observational tests of inflation with a field coupled to the Einstein
tensor was presented in \cite{Tsujikawa:2012mk}, while in
\cite{Koutsoumbas:2013boa} it was shown that the derivative
coupling to gravity provided a natural mechanism to suppress the
overproduction of heavy particles after inflation. Particle
production after the end of inflation in the presence of
derivative couplings was also discussed in \cite{Sadjadi:2012zp}.

This interesting cosmological behavior of the derivative
coupling of a scalar field to the Einstein tensor stems from the
fact that this term introduces a scale in the theory and
effectively acts as a
 cosmological constant \cite{Sushkov:2009hk}.  As is well
 known,  the presence of the cosmological constant alters the local
 properties of spacetime, allowing the evasion of the no-hair
 theorems. Hairy black hole solutions were found
 in the presence of a cosmological constant. Therefore, it is natural to ask
 whether black hole solutions exist in gravity theories with
 derivative couplings in the absence of a cosmological constant.

  In the case of a positive cosmological constant with a
minimally coupled scalar field with a self-interaction potential,
black hole solutions were found in~\cite{Zloshchastiev:2004ny}, and
a numerical solution, albeit an unstable one, was presented in~\cite{Torii:1998ir}. If the scalar field is non-minimally coupled,
a solution exists with a quartic self-interaction
potential~\cite{Martinez:2002ru}. However, this solution was also shown to be
unstable~\cite{Harper:2003wt,Dotti:2007cp}. In the case of a
negative cosmological constant, stable solutions were found
numerically for spherical geometries~\cite{Torii:2001pg,
Winstanley:2002jt}, and an exact solution in asymptotically AdS
space with hyperbolic geometry was presented
in~\cite{Martinez:2004nb}, and later generalized to include
charge~\cite{Martinez:2005di}, and further generalized to
non-conformal solutions \cite{Kolyvaris:2009pc}. Also, an  exact
solution of a charged C-metric conformally coupled to a scalar
field was presented in~\cite{Charmousis:2009cm, Anabalon:2009qt}.
Further hairy solutions in the presence of a cosmological constant
were reported in
\cite{Anabalon:2012ta,Anabalon:2012ih,Anabalon:2012tu,Bardoux:2012tr}
with various properties.

Black hole solutions in the general Horndeski theory are not
known. In theories resulting from the
truncation of higher-dimensional theories, which in four
dimensions give second-order field equations, black hole solutions were discussed
in \cite{Charmousis:2012dw}. In a gravity model with a scalar
field coupled to the Einstein tensor, an instability was found outside
the horizon of a Reissner-Nordstr\"om black hole, by calculating the quasinormal
spectrum of scalar perturbations
\cite{Chen:2010qf}. It was shown  that for higher angular momentum quantum numbers
and derivative coupling constant larger than a certain critical
value, the effective potential develops a negative gap near the
black hole horizon. This indicates that a
 phase transition  to a hairy black hole configuration can occur.

In our previous work \cite{Kolyvaris:2011fk}, we investigated this
effect in detail. We considered a gravity model consisting of an
electromagnetic field and a scalar field coupled to the Einstein
tensor with vanishing cosmological constant. We showed that a
Reissner-Nordstr\"om black hole undergoes a second-order phase
transition  to a hairy black hole configuration of generally
anisotropic hair at a certain critical temperature. Using perturbation theory,
we  calculated analytically the properties of the  hairy black hole
configuration near the critical temperature and showed that it is
energetically favorable over the corresponding
Reissner-Nordstr\"om black hole. Spherically symmetric black hole
solutions were also discussed in \cite{Germani:2011bc}.

The recent advances in holography, and in particular the
application of the  gauge/gravity duality to condensed matter
systems (for a review, see \cite{Hartnoll:2009sz}) has revived the
interest on the dynamics of a scalar field outside a black hole
horizon. The transition from a metallic state to a superconducting
state in a strongly-coupled system can be described by its dual weakly-coupled gravitational system using the AdS/CFT correspondence \cite{Maldacena:1997re}.
The simplest holographic superconductor model
\cite{Hartnoll:2008vx} is described by an Einstein-Maxwell-scalar
field theory with a negative cosmological constant. At high enough
temperatures the black hole-scalar system is decoupled and the
black hole is stable. The boundary gauge theory describes a metallic
state. When the temperature is lowered, the black hole becomes
unstable and a new black hole configuration with scalar
hair forms. The dual gauge theory describes a superconducting state (for a
review, see \cite{Horowitz:2010gk} and references therein). An
exact gravity dual of a gapless holographic superconductor was discussed
in \cite{Koutsoumbas:2009pa}.

The dynamics of a holographic superconductor depends crucially on
the behavior of the scalar field near the horizon of the black hole.
A model was presented in \cite{Gubser:2005ih,Gubser:2008px}
consisting of an electromagnetic field and a scalar field
minimally coupled to gravity in the presence of a negative
cosmological constant. It was shown that the effective mass of the
scalar field becomes negative for  large values of the charge of
the background Reissner-Nordstr\"om black hole, thus breaking an Abelian gauge symmetry outside the horizon of the
Reissner-Nordstr\"om black hole. In the language of gauge/gravity duality, this corresponds to the formation of a condensate  in the
boundary gauge theory, which below a certain critical temperature triggers the
transition from a metallic state to a superconducting one. A heuristic
way to explain this
behavior of the scalar field outside the black hole horizon was presented in \cite{Gubser:2008px}. If
the scalar particle is  highly charged,  then its gravitational
attraction to the black hole is overcome by its electrostatic
repulsion, and because the space has a boundary (being
asymptotically AdS), it reflects back and condenses outside the
black hole horizon. A generalization to higher dimensions and to
other horizon topologies was presented in
\cite{FernandezGracia:2009em}.

In this work, our main motivation is to study possible instabilities of a black hole near its horizon in asymptotically flat space (vanishing cosmological constant). We consider a gravitational system consisting of
an electromagnetic field and a charged scalar field which, together with
the standard minimal coupling to gravity, also couples to the Einstein
tensor. As it was discussed in \cite{Gubser:2008px}, in the
absence of the derivative coupling to the Einstein tensor, we cannot have a ``geometrical" breaking
of the Abelian symmetry near the black hole horizon, since the space is asymptotically flat.

By turning on the derivative coupling to the Einstein tensor, and solving
the resulting dynamical system of Einstein-Maxwell-scalar field equations, we will show that
 there is a critical temperature at which a phase transition  to a
 hairy black hole configuration occurs. The dimensionful coupling constant of the derivative coupling to the Einstein tensor provides the scale for the confining potential, which is an effect similar to the AdS radius provided by the cosmological constant. We argue that the new hairy black hole configuration results from the breaking of the Abelian symmetry near the horizon by curvature effects. We obtain a hairy black hole
 which is spherically symmetric.

Our discussion is organized as follows. In Section~\ref{sect1}, we
set up the theory and derive the field equations. In
Section~\ref{sect2}, we solve the field equations perturbatively. We discuss the solution at leading as well as next-to-leading order. We present both analytic and numerical results.
Finally in
Section~\ref{sect5}, we conclude.

\section{The field equations}
\label{sect1}

Our system consists of a $U(1)$ gauge potential $A_\mu$, and a scalar field $\varphi$ of mass $m$ and charge $q$, in a dynamical gravitational background with vanishing cosmological constant. Thus, spacetime is asymptotically flat. The action is
\be\label{EGBscalaraction}
   I=\int  d^4x\sqrt{-g}\left[ \frac{R}{16\pi G}-\fr{1}{4}F_{\mu\nu}F^{\mu\nu}-(g^{\mu\nu}-\k
   G^{\mu\nu})D_\mu\varphi (D_\nu\varphi)^{*} - m^2|\varphi|^2 \right]~,
\ee
where $D_\mu = \nabla_\mu - i q A_\mu$, and $F_{\mu\nu} = \partial_\mu A_\nu - \partial_\nu A_\mu$ is the field strength.
The action \eqref{EGBscalaraction} contains a derivative coupling of the scalar field to Einstein tensor with coupling constant $\k$ of
dimension length squared. For convenience, we introduce the notation
 \be
\Phi_{\mu\nu} \equiv D_{\mu}\varphi (D_{\nu}\varphi)^*~,\ \
\Phi \equiv g^{\mu\nu}\Phi_{\mu\nu}~.
 \ee
The field equations resulting from the variation of the action
\eqref{EGBscalaraction} are the Einstein equations
 \bea G_{\mu\nu} = 8\pi G T_{\mu\nu} \
\ , \ \ \ \ T_{\mu\nu} = T_{\mu\nu}^{(\varphi)} +
T_{\mu\nu}^{(EM)} - \k\Theta_{\mu\nu}~, \label{einst}\eea where,
\bea
T_{\mu\nu}^{(\varphi)} & = &   \Phi_{\mu\nu} + \Phi_{\nu\mu} - g_{\mu\nu}(g^{ab}\Phi_{ab} + m^2 |\varphi|^2)~, \\
T_{\mu\nu}^{(EM)} & = & F_{\mu}^{\phantom{\mu} \alpha} F_{\nu
\alpha} - \fr{1}{4} g_{\mu\nu} F_{\alpha\beta}F^{\alpha\beta}~,
\eea\bea
\Theta_{\mu\nu}  = & -& g_{\mu\nu} R^{ab}\Phi_{ab} + R_{\nu}^{\phantom{\nu}a}(\Phi_{\mu a} + \Phi_{a\mu}) + R_{\mu}^{\phantom{\mu}a} (\Phi_{a\nu} + \Phi_{\nu a})  - \fr{1}{2} R (\Phi_{\mu\nu} + \Phi_{\nu\mu}) \nn\\
& - & G_{\mu\nu}\Phi - \fr{1}{2}\nabla^a\nabla_\mu(\Phi_{a\nu} + \Phi_{\nu a}) - \fr{1}{2}\nabla^a\nabla_\nu(\Phi_{\mu a} + \Phi_{a\mu}) + \fr{1}{2}\Box (\Phi_{\mu\nu} + \Phi_{\mu\nu}) \nn \\
& + & \fr{1}{2}g_{\mu\nu} \nabla_a\nabla_b (\Phi^{ab} + \Phi^{ba})
+ \fr{1}{2}(\nabla_\mu\nabla_\nu + \nabla_\nu\nabla_\mu) \Phi -
g_{\mu\nu}\Box\Phi~, \label{theta} \eea
the Klein-Gordon equation \be (\de_\mu-i q A_\mu) \left[
\sqrt{-g}(g^{\mu\nu} - \k G^{\mu\nu})(\de_\nu - i q A_\nu)\varphi \right] =
\sqrt{-g} m^2 \varphi~, \label{glgord} \ee
and the Maxwell equations \be \nabla_\nu F^{\mu\nu} +
(g^{\mu\nu} - \k G^{\mu\nu}) \left[ 2 q^2 A_\nu |\varphi|^2 + i q
(\varphi^*\nabla_\nu\varphi - \varphi\nabla_\nu\varphi^*)\right]
=0~. \label{max}\ee Our goal is to find spherically
symmetric solutions of the coupled system of
Einstein-Maxwell-scalar equations (\ref{einst}),
(\ref{glgord}) and (\ref{max}). Apart from the complexity of the
$\Theta_{\mu\nu}$ contribution (\ref{theta}) to the energy-momentum tensor
$T_{\mu\nu}$, there is another  technical difficulty in solving
the system of field  equations. The presence of the dimensionful
coupling $\k$ does not allow the scalar field to be conformally
coupled to gravity. In many of the existing  hairy black hole
solutions, the conformal symmetry in the scalar sector helps to
find exact  solutions
\cite{Martinez:2004nb,Martinez:2005di,Charmousis:2009cm}.

For a spherically symmetric solution, consider the metric \emph{ansatz}
 \be \frac{ds^2}{\mu^2} = -e^{-\alpha (z)}
dt^2 + \frac{l(z) e^{\alpha(z)}}{z^2} \left[ \frac{dz^2}{z^2} +
d\Omega^2 \right] \label{papametr}~, \ee where $\mu$ is an arbitrary scale and all metric functions depend only on the radial coordinate $z$ (no dependence on
the angles $\Omega = (\theta,\phi)$, or time $t$). We will place the horizon at $z=1$, and choose coordinates so that $le^\alpha$ remains finite at the horizon.
Furthermore, we assume asymptotic flatness as $z\to 0$ (so $\alpha\to 0$, and $l\to
1$). The arbitrary scale $\mu$ can be thought
of as the position of the horizon, by switching the radial coordinate to $r$ defined by $z = \frac{\mu}{r}$. After rescaling $t\to t/\mu$, the metric no longer has an explicit dependence on the  parameter $\mu$ (depending on it only through the metric functions $\alpha$ and $l$),
and the horizon is placed at $r=\mu$. In calculations, we may
safely ignore $\mu$, setting $\mu =1$, but we need to restore it
in dimensionful quantities (e.g., $\kappa \to \kappa/\mu^2$).

The temperature of the black hole is (recall $le^\alpha$ is finite at the horizon, by our choice of coordinates)
\be\label{eqT} T =
\frac{1}{2\pi \mu l(1) e^\alpha (1)} \sqrt{\frac{l''(1)}{2}}~. \ee
Using the metric \emph{ansatz} \eqref{papametr}, the field equations reduce to the following equations.

The Einstein
equations \eqref{einst} are conveniently written as \be R_{\mu}^{\nu} = 8\pi G \left[ T_{\mu}^{\nu} -
\frac{1}{2} \delta_{\mu}^{\nu} T_\alpha^\alpha \right]~.
\label{einmetr}\ee
Only the diagonal components are non-vanishing, and the two angular components are equal to each other by spherical symmetry. Therefore, there are three independent equations with corresponding components of the Ricci tensor,
\bea R_t^t &=& -\frac{z^4e^{-\alpha}}{4\mu^2 l} \left[ \frac{l'}{l}\alpha' + 2\alpha'' \right]~, \nonumber\\
R_z^z &=& \frac{z^4 e^{-\alpha}}{4\mu^2 l} \left[ -4 \frac{{l'}^2}{l^2} + \frac{l'}{l}\left( \frac{4}{z}+\alpha'
\right) + 4 \frac{l''}{l}+2 {\alpha'}^2 + 2\alpha'' \right]~, \nonumber\\
R_\theta^\theta &=& R_\phi^\phi =  \frac{z^4
e^{-\alpha}}{4\mu^2 l} \left[  -\frac{{l'}^2}{l^2} +\frac{l'}{l} \left(
-\frac{2}{z}+\alpha' \right) +2\frac{l''}{l}+2\alpha'' \right]~.
\eea The various contributions to the stress-energy tensor  are \be T_t^{(EM)t} =
T_z^{(EM)z} = - T_\theta^{(EM)\theta} = - T_\phi^{(EM)\phi} =
\frac{z^4}{2\mu^2 l} {A_t'}^2~, \ee
\bea \mu^2 T_t^{(\varphi)t} &=& -\frac{1}{4} \left[ m^2 + q^2 e^\alpha A_t^2 \right] \varphi^2 - \frac{z^4}{4 l} e^{-\alpha} {\varphi'}^2~, \nonumber\\
\mu^2 T_z^{(\varphi)z} &=& -\frac{1}{4} \left[ m^2 - q^2 e^\alpha A_t^2 \right] \varphi^2 + \frac{z^4}{4 l} e^{-\alpha} {\varphi'}^2~, \nonumber\\
\mu^2 T_\theta^{(\varphi)\theta} &=& \mu^2 T_\phi^{(\varphi)\phi}
=-\frac{1}{4} \left[ m^2 - q^2 e^\alpha A_t^2 \right] \varphi^2 -
\frac{z^4}{4 l} e^{-\alpha} {\varphi'}^2 ~.\eea and \bea
\mu^2\Theta_t^t &=& -\frac{q^2z^4}{16 l^3} A_t^2\varphi^2 \left[
-3{l'}^2 +2ll'\alpha' +4ll'' + l^2({\alpha'}^2 + 4\alpha'')
\right]
 -\frac{z^6e^{-2\alpha}}{16 l^4}
\varphi' \Big[ 7z^2 {l'}^2 \varphi' \nonumber\\
& & + (24-24 z\alpha' +3z^2 {\alpha'}^2 -4z^2 \alpha'') l^2
\varphi' - 8z(-2+z\alpha') l^2 \varphi''
-4z^2 ll''\varphi'~ \nonumber\\
& & +2z (3(-4+z\alpha')\varphi' -4z\varphi'')ll' \Big]~, \nonumber\\
\mu^2\Theta_z^z &=& -\frac{q^2z^3}{16 l^3} A_t^2\varphi \left[
z{l'}^2\varphi +4ll'((9-1+z\alpha')\varphi +2z\varphi') \right]
 +\frac{q^2z^3}{2 l^2} A_tA_t'\varphi^2 \left[ zl'+(-2+z\alpha')l \right] \nonumber\\
&& \frac{z^6e^{-2\alpha}}{16 l^4}
{\varphi'}^2 \left[ -12zll'+3z^2 {l'}^2 + (8-3z^2 {\alpha'}^2)l^2 \right]~,  \nonumber\\
\mu^2 \Theta_\theta^\theta &=& \mu^2\Theta_\phi^\phi = \frac{q^2z^3 }{16 l^3} A_t^2 \Big[ -2z{l'}^2 \varphi^2 +2l(l'+zl'')\varphi^2 +l^2 (8z{\varphi'}^2 + (4\alpha' + z{\alpha'}^2 +4z\alpha'')) \varphi^2 \nonumber\\
&& \quad\quad+8 ((1+z\alpha')\varphi'+z\varphi'')\varphi \Big] + \frac{q^2 z^4}{2 l} {A_t'}^2 \varphi^2 \nonumber\\
&& \quad\quad + \frac{q^2 z^3}{2 l} A_t \varphi \left[ 4zA_t'\varphi' + ((1+z\alpha') A_t' + zA_t'')\varphi \right]+ \frac{z^7 e^{-2\alpha}}{4 l^3} \left[  -2l +zl' \right]\varphi'\varphi'' \nonumber\\
&& \quad\quad + \frac{z^6 e^{-2\alpha}}{16 l^4} {\varphi'}^2
\left[ -4z^2 {l'}^2 + (-16+4z\alpha'+z^2 {\alpha'}^2) l^2 + 2z
((7-z\alpha')l'+zl'')l \right]~. \nonumber \\ \eea The
Klein-Gordon equation \eqref{glgord} reads \be\label{eq28}
\frac{z^4 e^{-\alpha}}{l^{3/2}}\left(\sqrt{l}\mathcal{A} \varphi' \right)'
- \mu^2 m_{\text{eff}}^2 \varphi = 0~, \ee where
\bea \mathcal{A} & = & 1 + \frac{\kappa z^3 e^{-\alpha}}{4\mu^2l} \left[ \frac{4l'}{l}-\frac{z{l'}^2}{l^2} +z {\alpha'}^2 \right]~, \nonumber\\
m_{\text{eff}}^2  &=& m^2 - q^2 e^\alpha A_t^2
\left[ 1 - \frac{\kappa z^4 e^{-\alpha}}{4\mu^2 l} \left(
-\frac{3{l'}^2}{l^2} + 2\alpha' \frac{l'}{l} + \frac{4l''}{l} +
{\alpha'}^2 +4\alpha'') \right) \right]. \label{beta}\eea Finally, the
Maxwell equations reduce to Gauss's Law, \be A_t'' + \left[
\frac{l'}{2l} + \alpha' \right] A_t' + \frac{q^2}{2} \left[
\frac{e^\alpha l}{z^4} - \frac{\kappa}{4\mu^2}\left(
-\frac{3{l'}^2}{l^2} +2\alpha'\frac{l'}{l}+ \frac{4l''}{l} +
{\alpha'}^2 + 4\alpha'') \right) \right]\varphi^2 A_t = 0.
\label{maxmetr}\ee The above system of non-linear equations need
to be solved for the functions $\alpha, l , A_t, \varphi$. We
shall set $16\pi G = 1$.

\section{Perturbative solution}
\label{sect2}

To solve the coupled non-linear system of equations
(\ref{einmetr}), (\ref{eq28}) and (\ref{maxmetr}), we use
perturbation theory. We expand the metric functions, and the
scalar potential  for small values of the scalar field $\varphi$. Introducing the bookkeeping parameter $\epsilon$, we have
\bea \alpha &=& \alpha_0 + \epsilon\alpha_1 + \epsilon^2 \alpha_2 + \dots \nonumber\\
l &=& l_0 + \epsilon l_1 + \epsilon^2 l_2 + \dots \nonumber\\
A_t &=& A_{t0} + \epsilon A_{t1} + \epsilon^2 A_{t2} + \dots \nonumber\\
\varphi &=&  \epsilon \varphi_0 + \epsilon^2 \varphi_1 + \epsilon^3 \varphi_2 + \dots ~.\eea

\subsection{Leading order}

At zeroth order, we
obtain the system
\bea \frac{l_0'}{l_0}\alpha_0' + 2\alpha_0'' + e^{\alpha_0} {A_{t0}'}^2 &=& 0~, \nonumber\\
 -4z \frac{{l_0'}^2}{l_0^2} + \frac{l_0'}{l_0}(4+z\alpha_0') + 4z\frac{l_0''}{l_0}+2z ({\alpha_0'}^2 +
 \alpha_0'')-  z e^{\alpha_0} {A_{t0}'}^2 &=& 0~,\nonumber\\
 -z\frac{{l_0'}^2}{l_0^2} +\frac{l_0'}{l_0} (-2+z\alpha_0') +2z\frac{l_0''}{l_0}+2z\alpha_0''  +   z e^{\alpha_0} {A_{t0}'}^2 &=& 0~,\nonumber\\
 A_{t0}'' + \left[ \frac{l_0'}{2l_0} + \alpha_0' \right] A_{t0}' &=&
 0~, \nonumber\\
\frac{z^4 e^{-\alpha_0}}{l_0^{3/2}}\left(\sqrt{l_0}\mathcal{A}_0 \varphi_0' \right)'
- \mu^2 m_{\text{eff},0}^2 \varphi_0 &=& 0~,\label{eq20}\eea
where
\bea \mathcal{A}_0 & = & 1 + \frac{\kappa z^3 e^{-\alpha_0}}{4\mu^2l} \left[ \frac{4l_0'}{l_0}-\frac{z{l_0'}^2}{l_0^2} +z {\alpha_0'}^2 \right]~, \nonumber\\
m_{\text{eff},0}^2  &=& m^2 - q^2 e^{\alpha_0} A_{t0}^2
\left[ 1 - \frac{\kappa z^4 e^{-\alpha_0}}{4\mu^2 l_0} \left(
-\frac{3{l_0'}^2}{l_0^2} + 2\alpha_0' \frac{l_0'}{l_0} + \frac{4l_0''}{l_0} +
{\alpha_0'}^2 +4\alpha_0'') \right) \right]~. \label{beta0}\eea
If the scalar field vanishes (no hair), the Klein-Gordon equation is trivially satisfied, and the rest of the equations form a system of Einstein-Maxwell field equations whose
solution is the Reissner-Nordstr\"om black hole written in
Papapetrou coordinates, \be\label{eq22} \alpha_0 = \ln \frac{(1+2z\coth B +
z^2)^2}{l_0} \ \ , \ \ \ \ l_0 = (1-z^2)^2 \ \ , \ \ \ A_{t0} =
2e^{-B} - \frac{4z}{(1+z^2) \sinh B +2z\cosh B }. \ee The mass and
the charge are found from the $\mathcal{O} (z)$ terms in $\alpha$
and $A_t$. We obtain, respectively, \be
\frac{M}{\mu} = 8\coth B \ \ , \ \ \ \ \frac{Q}{\mu} =
\frac{4}{\sinh B}~. \label{ratio}\ee
Notice that the ratio $\frac{Q}{M} = \frac{1}{2\cosh B}$ is solely a function of $B$. As $B\to\infty$, $\frac{Q}{M} \to 0$, and we obtain the Schwarzschild black hole. For $B=0$, the ratio attains its maximum value, $\frac{Q}{M} = \frac{1}{2}$, corresponding to the extremal Reissner-Nordstr\"om black hole.

The temperature is found from
(\ref{eqT}) to be \be\label{eqT0} T = T_0 = \frac{e^{-2B} \sinh^2
B}{4\pi\mu}~. \ee
For the Schwarzschild black hole ($B\to\infty$), $T = \frac{1}{16\pi \mu}$, whereas for the extremal Reissner-Nordstr\"om black hole ($B =0$), $T=0$, as expected.


For a hairy solution, we need a non-vanishing scalar field. At zeroth order, we must have $\varphi_0 \ne 0$. Therefore, additionally, we must find the solution to the zeroth-order Klein-Gordon equation (last equation in \eqref{eq20}; the first four equations are independent of $\varphi_0$). This constrains the two-parameter family of solutions we just obtained (Reissner-Nordstr\"om black holes parametrized by $(M, Q)$, or equivalently $(\mu, B)$).

Using the  Reissner-Nordstr\"om metric functions \eqref{eq22}, the functions \eqref{beta0} entering the zeroth-order Klein-Gordon equation read
\bea\label{eq35} \mathcal{A}_0 & = & 1 + \frac{4\kappa}{\mu^2} \frac{ z^4 \sinh^2B}{[(1+z^2)\sinh B +2z\cosh B]^4}~, \nonumber\\
m_{\text{eff},0}^2  &=&  m^2 - 4q^2 e^{-2B} \frac{(1-z)^2}{ (1+z)^2}
- \frac{16q^2\kappa e^{-2B}\sinh^2 B}{\mu^2} \frac{z^4
(1-z)^2}{(1+z)^2 [(1+z^2)\sinh B +2z\cosh B]^4 }~. \eea
Notice that for $\k \ge 0$, we always have $\mathcal{A}_0 > 0$. Thus, $m_{\text{eff},0}^2$ provides an effective potential which determines the existence of solutions to the scalar equation (``bound states"). For a solution, we need $m_{\text{eff},0}^2 <0$ within a finite interval outside the horizon. Evidently, for sufficiently large charge of the scalar $q$, we have $m_{\text{eff},0}^2 <0$ in a long enough interval that guaranties the existence of a solution, implying a potential instability of the black hole. In the absense of the derivative coupling of the scalar field to the Einstein tensor ($\k =0$), this interval necessarily includes infinity ($z=0$), attaining the asymptotic value $m_{\text{eff},0}^2 = \nu^2 <0$, where
\be\label{eqnu} \nu^2 \equiv m^2 - 4q^2 e^{-2B}~. \ee
Consequently, a scalar particle flying away from the horizon (due to a repulsive electric force which exceeds the gravitational attraction) travels all the way to infinity and can never condense \cite{Gubser:2008px}. On the other hand, no instability occurs for $\nu^2 \ge 0$. This is in accord with the no-hair theorem in asymptotically flat spaces.

The above picture changes if $\k > 0$~\footnote{If
the coupling constant $\kappa$ is negative, then the system of
Einstein-Maxwell-Klein-Gordon equations \eqref{einmetr},
\eqref{eq28}, and \eqref{maxmetr}, is unstable and no solutions can
be found. We reached the same conclusion in our
previous work \cite{Kolyvaris:2011fk}. The sign of the
coupling constant $\k$ was also discussed in \cite{Germani:2010gm}, where it was
claimed that a negative $\kappa$ introduces ghosts into the
theory. Finally, in \cite{Koutsoumbas:2013boa} a very small window
of negative $\kappa$ was shown to be allowed.}. While for $\nu^2 < 0$ we still have no condensation, for $\nu^2 >0$ the additional terms proportional to $\k$ introduce a potential well which can trap scalar particles.  For large enough values of the coupling constant $\k$,  this well is deep enough for the scalar particles to condense, as we will show. Thus, the no-hair theorem is evaded and hair forms.
\begin{figure}[t]
\begin{center}
\includegraphics[scale=1.,angle=0]{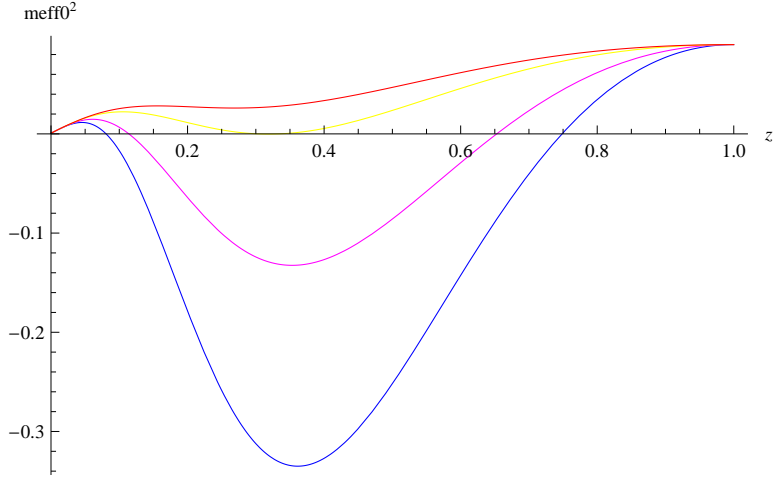}
\end{center}
\caption{The effective mass $m_{eff}^2$ \eqref{eq35} as a function
of the radial coordinate $z$ outside the horizon (located at $z=1$)
for coupling constant (top to bottom) $\kappa= 100,\ 170,\ 500,\
1000$, and mass and charge, respectively, of the scalar field (black
hole), $m=0.3,\ q=0.5$ ($M=3.2,\ Q=1.0$).} \label{meff1}
\end{figure}

%

In figure \ref{meff1}, we plot $m_{\text{eff},0}^2$ as a function of the radial coordinate $z$ for various (indicative) values of the parameters on which it depends, with $\nu^2 >0$. We chose $m = 0.3$ and $q=0.5$ for the scalar field. We also fixed the mass and charge of the Reissner-Nordstr\"om black hole to $M=3.2$ and $Q=1.0$, respectively. The four curves (top to bottom) correspond to the choices $\kappa = 100,\ 170,\ 500,\ 1000$, respectively. One can see the potential well forming in all cases, however for $\k = 100, 170$, the curve is above the axis, and the wave equation has no solutoin. For $\k = 500,\ 1000$, the potential well dips into negative values. This is a necessary but not sufficient condition for the existence of solutions. For the chosen values of mass and charge of the particle, a solution can be found corresponding to potential wells of the form depicted in fig.\ \ref{meff1} for $\k = 1000$, and mass and charge of the black hole near the chosen values.


Next, we discuss the numerical solution of the zeroth-order wave equation (last equation in \eqref{eq20} with $\mathcal{A}_0$ and $m_{\text{eff},0}^2$ given in \eqref{eq35}). It is convenient to isolate the asymptotic behavior of the scalar field. As $z\to 0$ ($r = \frac{\mu}{z} \to \infty$), we obtain
\be \varphi_0
(z) \sim z^{1+2\mu\xi} e^{-\fr{\mu\nu}{z}} \sim \frac{1}{r^{1+2\mu\xi} } e^{-\nu r} \ \ ,
\ \ \ \ \xi \equiv \fr{\nu}{\sinh B} - \frac{m^2 }{ \nu} \label{sca1}\ee
where $\nu$ is defined in \eqref{eqnu}. Notice that the field decays exponentially as $r\to\infty$, to be contrasted with the standard power law behavior is AdS space (in which hairy solutions have been shown to exist). Thus, if a solution exists in our case, the condensation occurs in the vicinity of the horizon and decays rapidly away from it.

Defining
\be \varphi_0
(z) = z^{1+2\mu\xi} e^{-\fr{\mu\nu}{z}} \chi(z)~, \ee
we will solve the wave equation numerically for $\chi$. This function is regular at $z=0$, and is arbitrarily normalized, since it obeys a linear equation. One may choose $\chi (0) = 1$ as one of the boundary conditions. The other boundary condition is determined from the analytic solution to the wave equation expanded around $z=0$ (which is a regular point for $\chi$). We obtain
\be \chi (z) = 1+ \frac{ 1}{2\mu\nu}\left[ 2\mu\xi (1+2\mu\xi)-2 \mu ^2 \nu^2 \left( 5\nu^2+2  \nu^2 \coth ^2B - 8\nu^2 \coth B \right)+8 \mu^2 m^2  \left( 2\coth B-1  \right)\right] z + \mathcal{O} (z^2) ~, \ee
from which we one can read off $\chi'(0)$, to be used as a second boundary condition. Thus, for a given set of parameters of the scalar field $(m,q)$, one obtains a unique solution. However, these solutions are generally unacceptable, because they diverge at the horizon logarithmically ($\chi (z) \sim \ln (1-z)$, as $z\to 1$). If $\kappa =0$, all solutions diverge at the horizon, attesting to the validity of the no-hair theorem. For $\kappa >0$, there are pairs of black hole parameters $(M,Q)$ for which regular solutions exist in the entire interval $z \in (0,1]$. In the two-dimensional parameter space, they form a one-dimensional curve, separating the region of stability from the region of instability of a Reissner-Nordstr\"om black hole.

\begin{figure}
\begin{center}
\includegraphics[scale=1.,angle=0]{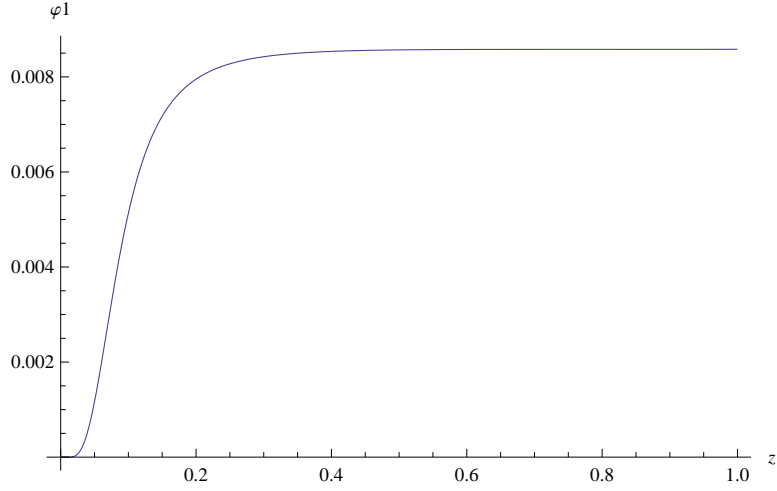}
\end{center}
\caption {The profile of the scalar field $\varphi_0$ outside
the horizon ($0< z\le 1$) for coupling constant $\k = 1000$,
and mass and charge, respectively, of the scalar field (black hole),
 $m=0.30,\ q=0.50$ ($M=3.187,\ Q=1.048$).}
\label{chi_phi1}
\end{figure}

Figure \ref{chi_phi1} depicts a sample profile of the scalar field
$\varphi_0$ obtained by numerically solving the wave equation as
outlined above. We chose the scalar field parameters $m=0.30$ and
$q=0.50$  A regular solution was found for the black hole parameters
$M=3.20$ and $Q=1.05$. The scalar field is concentrated within
$z\lesssim 0.2$, i.e., within about five times the horizon radius
(in radial coordinate $r = \frac{\mu}{z}$), beyond which it decays
exponentially ($\varphi_0 \sim e^{-\nu r}$, for $r \gtrsim 5\mu$).


\begin{figure}
\begin{center}
\includegraphics[scale=0.6,angle=-90]{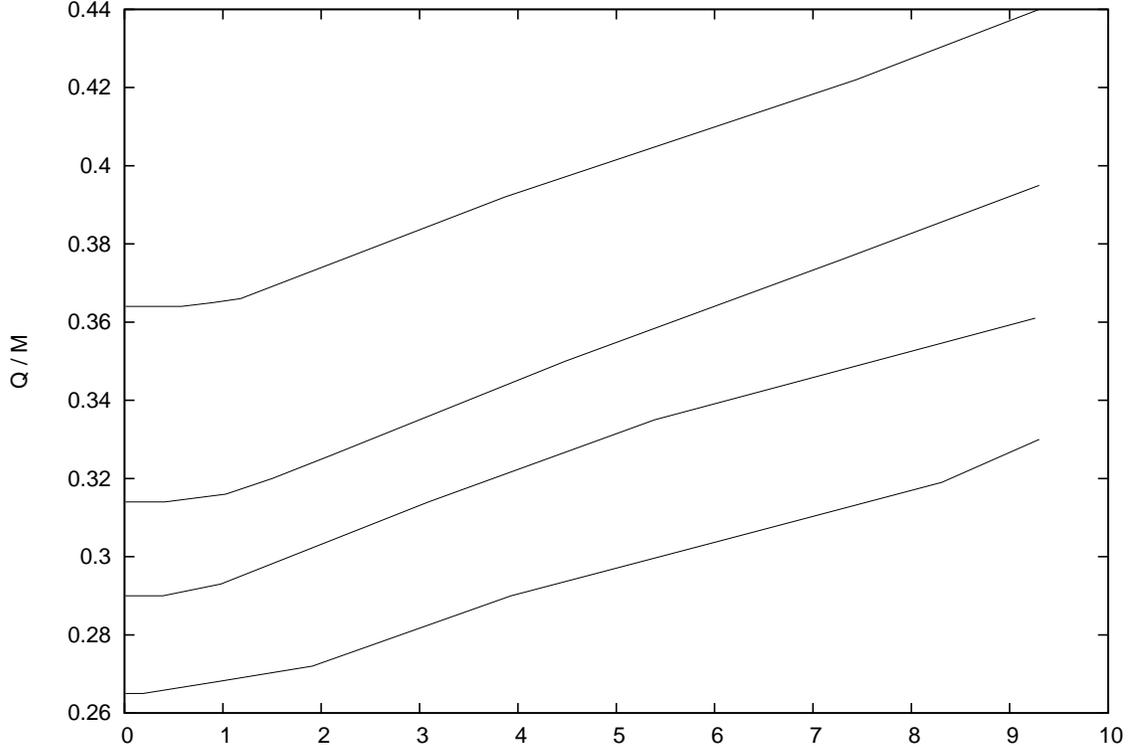}
\end{center}
\caption {Critical curves separating the region of instability (above each curve) from the region of stability (below the curve) for coupling constant (top to
bottom) $\kappa= 1000,\ 3000,\ 5000,\ 10000$, and mass and charge of
the scalar field $m=1.0,\ q=0.5,$ respectively. The horizontal axis is $M$ and the vertical axis is $Q/M$, where $M$ and $Q$ are the mass and charge of the black hole, respectively. } \label{QM}
\end{figure}

Finally in figure \ref{QM}, we plot critical lines for various values of the coupling constant $\kappa$ ($\kappa= 1000$, $3000$, $5000$, $10000$, from top to bottom). Below each critical line, the Reissner-Nordstr\"om black hole is stable, whereas above it, an instability arises.  We chose fixed values of the mass and charge of the scalar, $m=1.0,\ q=0.5,$ respectively. Notice that as $\kappa$ decreases, the region of instability shrinks toward the extremal value $Q/M = 0.5$ (zero temperature). This is expected, because the wave equation has no regular solutions for small $\k$ (for $\k =0$, such solutions are forbidden by the no-hair theorem).

\subsection{Next-to-leading order}

Having obtained the zeroth-order solution to the field equations, we now turn to the derivation of the next-to-leading order corrections to the metric, the electrostatic potential, and the scalar field. Since the scalar field enters the Einstein-Maxwell equations quadratically (through the stress-energy tensor), it follows that the $\mathcal{O} (\epsilon)$ corrections vanish ($\alpha_1 = l_1 = A_{t1} =0$). Therefore, the next-to-leading order corrections are given by the $\mathcal{O} (\epsilon^2)$ terms.

\begin{figure}
\begin{center}
\includegraphics[scale=0.9,angle=0]{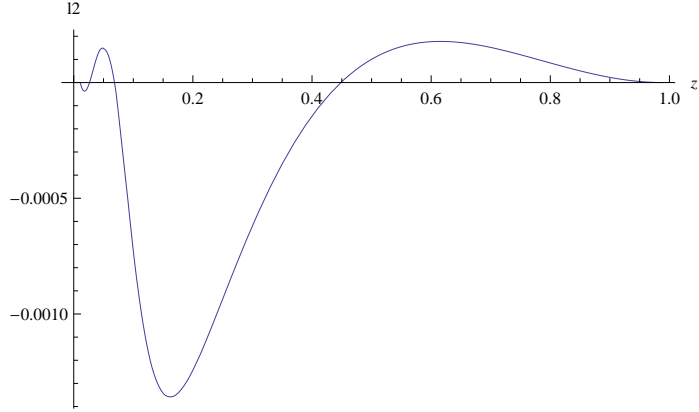}
\end{center}
\caption {Next-to-leading-order metric function $l_2(z)$ for choice of parameters as in figure \ref{chi_phi1}.} \label{ll2}
\end{figure}

From the sum
of the $tt$ and $\theta\theta$ components of the Einstein equations, we  obtain at $\mathcal{O} (\epsilon^2)$,
\be l_2''-\frac{1-5z^2}{z(1-z^2)} l_2'+
\frac{8z^2}{(1-z^2)^2} l_2 = \mathcal{L}~, \ee where the right-hand side consists of zeroth-order functions which have been already calculated,
\bea\label{eqL} \mathcal{L} &=& -\kappa \left[ q^2 e^{\alpha_0} l_0 A_{t0}^2\varphi_0 + z^3 e^{-\alpha_0} \left( -1 + \frac{zl_0'}{2l_0} \right) \varphi_0' \right] \varphi_0''\nonumber\\
&& + \kappa e^{-\alpha_0} \left[ - q^2 e^{2\alpha_0} l_0 A_{t0}^2
+ \frac{z^4 {l_0'}^2}{8l_0^2} + \frac{z^2}{4} ( 4 -2z\alpha_0' +
z^2 {\alpha_0'}^2 )
+ \frac{z^3}{4} \left( (-1+z\alpha_0') \frac{l_0'}{l_0} - z \frac{l_0''}{l_0} \right) \right] {\varphi_0'}^2\nonumber\\
&& -q^2\kappa e^{\alpha_0} l_0 A_{t0} \varphi_0\varphi_0' \left[ 4A_{t0}' + A_{t0} \left( -\frac{1}{z} + 2\alpha_0' + \frac{l_0'}{l_0} \right) \right]\nonumber\\
&& + \frac{e^{\alpha_0} l_0^2}{z^4} \left[ -m^2 + q^2 e^{\alpha_0}
A_{t0}^2 \right] \varphi_0^2  - \frac{q^2 \kappa
e^{\alpha_0}l_0}{8z} \varphi_0^2
\Big[ 8z {A_{t0}'}^2 \nonumber\\
&& + 8 A_{t0} \left( z \frac{l_0'}{l_0} A_{t0}' + (-1+2z\alpha_0') A_{t0}' + zA_{t0}'' \right) \nonumber\\
&& + A_{t0}^2\left( - z \frac{{l_0'}^2}{l_0^2} + 2
(-1+2z\alpha_0') \frac{l_0'}{l_0} +2 z \frac{l_0''}{l_0}
-4\alpha_0' + 4z{\alpha_0'}^2 +4z\alpha_0'' \right) \Big]~. \eea
Notice that $\mathcal{L}$ vanishes both at the horizon and at infinity (at
 $z=0,\ 1$). This equation can be solved for $l_2$ analytically. We
obtain \be l_2 (z) = - z^2(1-z^2)\int_z^1 dw
\frac{\mathcal{L}(w)}{2w} - (1-z^2)^2 \int_0^z dw
\frac{w\mathcal{L}(w)}{2(1-w^2)}~,\label{ll2sol}\ee where we fixed
the integration constants by demanding $l_2(0) =l_2'(1)=0$ ($l$ must
have a double zero at the horizon, $z=1$). It is plotted in fig.\ \ref{ll2} for the same values of the various parameters that were used for the plot of the zeroth-order scalar field (fig.\ \ref{chi_phi1}).

Next, we look at the $tt$ component of the Einstein equations together with Gauss's Law. At $\mathcal{O} (\epsilon^2)$, they form a system of coupled linear equations to be
solved for $\{ \alpha_2, A_{t2} \}$, \bea
\alpha_2'' + \frac{l_0'}{2l_0} \alpha_2' - \left[ \alpha_0'' + \frac{l_0'}{2l_0} \alpha_0'
\right] \alpha_2 +e^{\alpha_0} A_{t0}' A_{t2}' &=& \mathcal{P}~, \label{sa2}\\
A_{t2}'' + \left[ \frac{l_0'}{2l_0} + \alpha_0'\right] A_{t2}'  +
A_{t0}' \alpha_2' &=& \mathcal{Q}~, \label{la2}\eea where the right-hand sides consist of terms dependent on zeroth-order functions only,
\bea
\mathcal{P} &=& \frac{\kappa}{4\mu^2 l_0} z^3e^{-\alpha_0} \left[
\frac{zl_0'}{l_0} -2 + z\alpha_0' \right] \varphi_0'\varphi_0''
+ \frac{1}{8} {\varphi_0'}^2 + \frac{e^{\alpha_0} l_0}{8z^4} \left[ m^2 + e^{\alpha_0} q^2 A_{t0}^2 \right] \varphi_0^2 \nonumber\\
&& - \frac{\kappa z^2 e^{-\alpha_0}}{32\mu^2 l_0} \left[ \frac{7z^2{l_0'}^2}{l_0^2} +\frac{6zl_0'}{l_0} (-4 +z\alpha_0') -\frac{4z^2l_0''}{l_0} + 24 -24 z\alpha_0' + 3z^2 {\alpha_0'}^2 -4z^2 \alpha_0'' \right] {\varphi_0'}^2\nonumber\\
&& - \frac{q^2\kappa e^{\alpha_0} A_{t0}^2}{32} \left[ -3 \frac{{l_0'}^2}{l_0^2} + 2\frac{l_0'}{l_0} \alpha_0' + 4\frac{l_0''}{l_0} + {\alpha_0'}^2 + 4\alpha_0'' \right] \varphi_0^2 - \frac{l_2''}{l_0} + \frac{l_2'}{2l_0} \left[ \frac{3l_0'}{l_0} - \alpha_0' \right] \nonumber\\
&& + \frac{l_2}{4l_0} \left[ - 9 \frac{{l_0'}^2}{l_0^2} + 4 \frac{l_0'}{l_0} \alpha_0' + 8 \frac{l_0''}{l_0} \right]~,\nonumber\\
\mathcal{Q} &=& -\frac{  q^2 A_{t0}^2e^{\alpha_0} l_0}{2z^4} \varphi_0^2 +\frac{q^2\kappa A_{t0}^2}{8} \left[ -3 \frac{{l_0'}^2}{l_0^2} + 2\frac{l_0'}{l_0} \alpha_0' + 4\frac{l_0''}{l_0} + {\alpha_0'}^2 + 4\alpha_0'' \right] \varphi_0^2\nonumber\\
&& - \frac{A_{t0}' l_2'}{2l_0} + \frac{l_2}{l_0} \left[ A_{t0}'
\left( \frac{l_0'}{l_0} + \alpha_0' \right) + A_{t0}'' \right]~.
\eea Equation (\ref{la2}) is of first-order in $A_{t2}'$, and yields
\bea A_{t2}' (z) &=& \frac{4(1-z^2)\sinh B }{[(1+z^2)\sinh B + 2z\cosh
B]^2}\nonumber \\ && \times\left[ 4\sinh B \alpha_2 (z)+ C + \int_0^z
dw \frac{[(1+w^2)\sinh B +2w \cosh B]^2\mathcal{Q} (w)}{1-w^2}
\right]~,\label{a2t} \eea where $C$ is an arbitrary integration
constant. The correction to the electrostatic potential is deduced by integrating and using $A_{t2} (1) =0$,
\be A_{t2} (z) = -\int_z^1 dw A_{t2}' (w)~. \ee
It provides a correction to the charge $Q$ of the black hole,
\be\label{eq38} \frac{\delta Q}{\mu} =
-\epsilon^2 A_{t2}'(0) = -\frac{4\epsilon^2C}{\sinh B} = - C\epsilon^2 \frac{Q}{\mu} ~, \ee where we used $\alpha_2(0)=0$.

Then using
(\ref{a2t}),  equation (\ref{sa2}) becomes a second-order equation in $\alpha_2$,
\be \alpha_2'' -
\frac{2z}{1-z^2} \alpha_2' - \frac{8}{[(1+z^2)\sinh B + 2z\cosh
B]^2} \alpha_2= \mathcal{P}'~, \label{a2eqn}\ee where once again the right-hand side is a zeroth-order (known) function, \be
\mathcal{P}' = \mathcal{P} + \frac{16\sinh B }{[(1+z^2)\sinh B +
2z\cosh B]^3} \left[   C + \int_0^z dw \frac{[(1+w^2)\sinh B +2w
\cosh B]^2\mathcal{Q} (w)}{1-w^2} \right]~, \ee to be solved for
$\alpha_2$ subject to the boundary conditions $\alpha_2(0)=0$ and
$\alpha_2(1) < \infty$. This equation can be solved numerically.

The solution thus obtained involves an arbitrary parameter $C$. This parameter will be fixed by solving the scalar equation at $\mathcal{O} (\epsilon^2)$. The latter reads
%
\bea \label{eq44} \frac{z^4 e^{-\alpha_0}}{l_0^{3/2}} \left( \sqrt{l_0}
\mathcal{A}_0 \varphi_2' \right)' - \mu^2 m_{\text{eff}, 0}^2 \varphi_2
&=& \frac{z^4 e^{-\alpha_0} l_2}{2l_0^{5/2}} \left( \sqrt{l_0} \mathcal{A}_0
\varphi_0' \right)' - \frac{z^4 e^{-\alpha_0}}{l_0^{3/2}} \left(
\frac{1}{\sqrt{l_0}} \left( -\frac{\kappa z^3
e^{-\alpha_0}}{\mu^2}\mathcal{A}_1 + \frac{l_2}{2} \mathcal{A}_0
\right) \varphi_0' \right)' \nonumber
\\ &+& \mu^2 m_{\text{eff},1}^2 \varphi_0~, \eea where the right-hand side consists of known functions (zeroth-order as well as next-to leading order that we have already calculated),
\bea \mathcal{A}_1 &=&
\frac{l_2'}{l_0^2} \left[ -1 + \frac{zl_0'}{2l_0} \right] +
\frac{l_2}{4l_0} \left[ \frac{8l_0'}{l_0} - \frac{3z
{l_0'}^2}{l_0^2} + z{\alpha_0'}^2 \right] + \frac{z}{2}
\alpha_0'\alpha_2' + \alpha_2 \left[ \frac{l_0'}{l_0} -
\frac{z{l_0'}^2}{4l_0^2
} + \frac{z}{4} {\alpha_0'}^2 \right]~, \nonumber\\
m_{\text{eff},1}^2 &=& -m^2 \left[ \frac{l_2}{l_0}+\alpha_2\right]+ q^2 A_{t0} e^{\alpha_0} \left[ 2 A_{t2} +\frac{ l_2}{l_0} A_{t0} \right] \nonumber\\
&& + \frac{\kappa }{4\mu^2 } q^2 \frac{z^4}{l_0} A_{t0}A_{t2} \left[ -\frac{3{l_0'}^2}{l_0^2} + 2\alpha_0' \frac{l_0'}{l_0} + \frac{4l_0''}{l_0} + {\alpha_0'}^2 +4\alpha_0'' \right]\nonumber\\
&& + \frac{\kappa }{4\mu^2 } q^2 \frac{z^4}{l_0} A_{t0}^2 \left[ \alpha_2'' +\frac{\alpha_2'}{2} \left( \frac{l_0'}{l_0} + \alpha_0' \right) +\frac{\alpha_2}{4} \left( -\frac{3{l_0'}^2}{l_0^2} + 2\alpha_0' \frac{l_0'}{l_0} + \frac{4l_0''}{l_0} + {\alpha_0'}^2 +4\alpha_0'' \right) \right]\nonumber\\
&& +  \frac{\kappa }{4\mu^2 } q^2 \frac{z^4}{l_0} A_{t0}^2 \left[
\frac{l_2''}{l_0} + \frac{l_2'}{2l_0} \left( - \frac{3l_0'}{l_0} +
\alpha_0' \right) + \frac{l_2}{l_0} \left( -
\frac{3{l_0'}^2}{l_0^2} + \frac{l_0'}{l_0} \alpha_0' +
2\frac{l_0''}{l_0} \right) \right]~. \eea After multiplying both
sides of (\ref{eq44}) by $\frac{e^{\alpha_0} l_0^{3/2}}{z^4} \varphi_0$, and integrating
over the interval $[0,1]$,  the left-hand side vanishes on account of
(\ref{eq28}) and (\ref{eq35}). We deduce
\be 0 = \int_0^1 dz (1-z^2) \left[\left( \frac{l_2}{2l_0}\right)'
\mathcal{A}_0 \varphi_0\varphi_0'  - \frac{\kappa
e^{-\alpha_0}}{\mu^2 l_0} \mathcal{A}_1{\varphi_0'}^2 -
\mu^2\frac{e^{\alpha_0} l_0}{z^4} m_{\text{eff},1}^2 \varphi_0^2 \right]~. \label{int3} \ee
Despite appearances, this is a simple linear algebraic equation in the unknown parameter $C$. It is easily solved analytically for $C$.

For the choice of parameters in figure \ref{chi_phi1}, we obtain from \eqref{int3}, $C = 8.0$.

\begin{figure}
\begin{center}
\includegraphics[scale=0.9,angle=0]{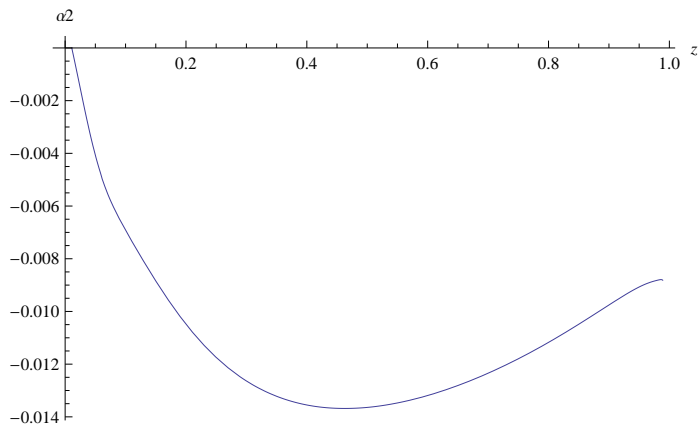}
\end{center}
\caption {Next-to-leading-order metric function $\alpha_2(z)$ for
choice of parameters as in figure \ref{chi_phi1}. In addition we
have chosen $\alpha_2^\pr(0)=-0.0000105.$ } \label{alpha2}
\end{figure}

Using this value, we can then solve \eqref{a2eqn} for $\alpha_2$ numerically. Using the boundary condition $\alpha_2 (0) = 0$ and fixing $\alpha_2'(0)$ arbitrarily, we obtain a solution which, in general, diverges logarithmically at the horizon. By varying the value of $\alpha_2'(0)$, we pinpoint the regular solution. It is depicted in figure \ref{alpha2}.

\begin{figure}
\begin{center}
\includegraphics[scale=0.9,angle=0]{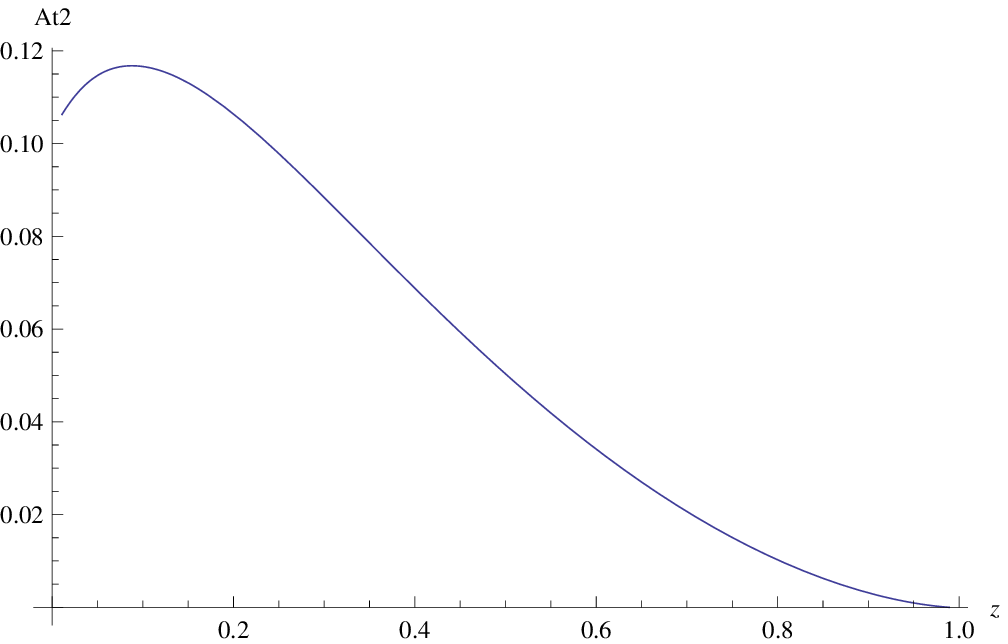}
\end{center}
\caption {Next-to-leading-order metric function $A_{t2}(z)$ for choice of parameters as in figure \ref{chi_phi1}.} \label{A2}
\end{figure}

Next, we use the numerical solution for $\alpha_2$ to determine the correction to the electrostatic potential $A_{t2}$. The result is depicted in figure \ref{A2}.

This completes the determination of all metric functions as well as the electrostatic potential at next-to-leading order.



%

The resulting hairy black hole has the following global properties. Its mass is
\be M = -2\mu \alpha'(0) = 8\mu \coth B -2\mu \epsilon^2 \alpha_2'(0) + \mathcal{O} (\epsilon^4)~, \ee
where $\alpha_2'(0)$ is determined from the solution of Eq.\ \eqref{a2eqn}, its charge is
\be Q = - \mu A_t'(0) =  \left[ 1 - C \epsilon^2 \right] \frac{4\mu}{\sinh B} + \mathcal{O} (\epsilon^4)~, \ee
where $C$ is determined from Eq.\ \eqref{int3}, and its temperature
can be found from  (\ref{eqT}) to be \be \frac{T}{T_0}
= 1 -\epsilon^2 \tau  + \mathcal{O} (\epsilon^4) \ \ , \ \ \ \ \tau = \left. \left(  \alpha_2 +
\frac{l_2}{l_0} - \frac{l_2''}{2l_0''} \right) \right|_{z=1} = \alpha_2(1)
+ \int_0^1 dz \frac{z\mathcal{L}(z)}{4(1-z^2)}~, \ee where $T_0$
is given by the zeroth-order approximation \eqref{eqT0}, $\alpha_2(1)$ is determined from the solution of Eq.\ \eqref{a2eqn}, and $\mathcal{L}$ is defined in \eqref{eqL}.

\section{Conclusion}
\label{sect5}

We considered a gravitating system without a cosmological constant term (therefore living in asymptotically flat space)
consisting of an electromagnetic field, and a massive charged
scalar field which had a derivative coupling to the Einstein tensor of strength $\k$. For small values of the
scalar field we solved the coupled system of
Einstein-Maxwell-scalar field equations perturbatively.

We found that for sufficiently large values of the coupling constant $\kappa$, an Abelian $U(1) $ gauge symmetry was broken
in the vicinity of the horizon of the black hole, leading to a new  black hole
configuration  with isotropic hair. A non-vanishing $\k$ allowed us to evade the no-hair theorem. The scale introduced by the dimensionful coupling constant $\k$ acted similarly to the scale of the negative cosmological constant in AdS space, producing a potential well that could trap scalar particles near the horizon. Unlike in AdS space, which yields a power law asymptotic behavior for the scalar field, in our case we showed that the scalar field decayed exponentially away from the horizon.

It would be interesting to calculate correlation functions in this background and explore the possibility of a correspondence to a gauge theory, similar to the gauge theory / gravity duality in AdS space. It would also be interesting to derive a full numerical solution of the field equations down to zero temperature, which would enable us to probe the ground state of the system. Work in this direction is in progress.


\acknowledgments{
T. K. acknowledges support from the Operational Program ``Education
and Lifelong Learning'' of the National Strategic Reference
Framework (NSRF) - Research Funding Program: Heracleitus II,
co-financed by the European Union (European Social Fund - ESF) and
Greek national funds.
}

\end{document}